%% file: Main-sigconf.tex
\PassOptionsToPackage{table}{xcolor}
\documentclass[sigconf,nonacm]{acmart}

\AtBeginDocument{%
  }

\setcopyright{acmlicensed}

\usepackage{enumitem}
\usepackage{multirow}
\usepackage{graphicx}
\usepackage{amsmath}
\usepackage{natbib}
\usepackage{dsfont}
\usepackage{wrapfig}

\usepackage[utf8]{inputenc} 
\usepackage[T1]{fontenc}    
\usepackage{hyperref}       
\usepackage{url}            
\usepackage{booktabs}       
\usepackage{amsfonts}       
\usepackage{nicefrac}       
\usepackage{microtype}      
\usepackage{xcolor}         
\usepackage{subcaption}

\usepackage[most]{tcolorbox}
\usepackage{xcolor}
 
\usepackage{amssymb}
\definecolor{promptbg}{RGB}{245,247,250}
\definecolor{promptframe}{RGB}{70,90,120}
\definecolor{prompttitle}{RGB}{35,45,70}

\newtcbtheorem[number within=section]{promptbox}{Box}{
  enhanced,
  breakable,
  colback=promptbg,
  colframe=promptframe,
  coltitle=white,
  colbacktitle=prompttitle,
  fonttitle=\bfseries\small,
  fontupper=\footnotesize\ttfamily,
  boxrule=0.4pt,
  arc=2pt,
  left=4pt,
  right=4pt,
  top=3pt,
  bottom=3pt,
}{box}

\usepackage{caption}

\usepackage[table]{xcolor}
\definecolor{lightgray}{gray}{0.93}
\begin{document}

\title{Recommendation as Generation: Unifying Personalized Video Generation and Recommendation at Industrial Scale}

\author{Yanhua Cheng$^{\ast}$\textsuperscript{1}, 
Bo Wang$^{\ast}$\textsuperscript{1}, 
Haotian Zhang$^{\ast}$$^{\S}$\textsuperscript{2},
Xinyuan Gao$^{\ast}$\textsuperscript{1}, 
Zhihui Yin\textsuperscript{1}, 
Ben Xue\textsuperscript{1}, 
Yongzhi Li\textsuperscript{1}, 
Jieting Xue\textsuperscript{1}, 
Ye Ma\textsuperscript{1}, 
Minquan Wang\textsuperscript{1}, 
Jiahui Li\textsuperscript{1}, 
Tianyu Xu\textsuperscript{1}, 
Zhiqiang Liu\textsuperscript{1}, 
Xiao Lin\textsuperscript{1}, \\ 
Shiyang Wen\textsuperscript{1}, 
Changcheng Li\textsuperscript{1}, 
Liu Liu\textsuperscript{2}, 
Quan Chen\textsuperscript{1}, 
Peng Jiang$^{\dag}$\textsuperscript{1}, 
Kun Gai\textsuperscript{1}}

\affiliation{
  \institution{
    \begin{minipage}[t]{0.35\textwidth}
      \centering
      \textsuperscript{1}Kuaishou Technology\\
      Beijing, China
    \end{minipage}
    \hspace{1cm} 
    \begin{minipage}[t]{0.35\textwidth}\centering
      \textsuperscript{2}Beihang University\\
      Beijing, China\end{minipage}
  }
  \country{}
}

\thanks{
  $^{\ast}$Equal contribution.\\
  $^{\S}$Work done during an internship at Kuaishou Technology.\\
  $^{\dag}$Corresponding author.
}

\renewcommand{\shortauthors}{Yanhua Cheng et al.}

\begin{abstract}

Traditional short-video recommendation systems match user interest to a fixed pool of pre-produced videos, which limits their ability to capture fine-grained and dynamic preferences. We propose \textbf{Recommendation-as-Generation} (\textbf{RaG}), a new paradigm that generates personalized videos on demand from inferred user interest. Our framework unifies generative recommendation and video generation through shared semantic IDs (SIDs), which disentangle video representation into content semantics and creative style semantics, enabling both fine-grained modeling of user interest and controllable generation of interest-aligned videos. We further develop \textbf{Video Generation Agents} (\textbf{VGAs}) that are conditioned on inferred SIDs to drive hierarchical planning and refinement for video creation, including visual composition, audio alignment, and artistic effect enhancement. To optimize the framework, we effectively introduce a synergistic cross-domain reward learning mechanism that jointly enforces interest alignment, user feedback, and video quality assessment. 

We deploy RaG~\footnote{Project page: \url{https://recommendation-as-generation.github.io/}} on an industrial-scale platform with over 400 million daily active users and evaluate it in a revenue-critical advertising scenario. Online A/B tests show up to \textbf{1.87\%} ad revenue improvement compared to a strong production GRM baseline, demonstrating its effectiveness in driving further revenue gains beyond generative recommendation.
Our results highlight a closed-loop generative system as a promising paradigm for integrating personalized video generation into recommendation. 

\end{abstract}



\keywords{Generative Recommendation, Personalized Video Generation, \\Agents, Semantic Quantization, Reward Learning}


\maketitle


\input{kdd/1Introduction}

\input{kdd/paradigm}
\input{kdd/3Method}
\input{kdd/Deployment}

\input{kdd/4Experiment}

\input{kdd/2RelatedWork}

\input{kdd/6Conclusion}

\bibliographystyle{plainnat}
\bibliography{sample-base}

\clearpage
\appendix
\input{AppendixText/Appendix_full}

\end{document}

%% file: kdd/1Introduction.tex
\section{Introduction}\label{sec:intro}

\begin{figure}[t]
        \centering
        \includegraphics[width=0.47\textwidth]{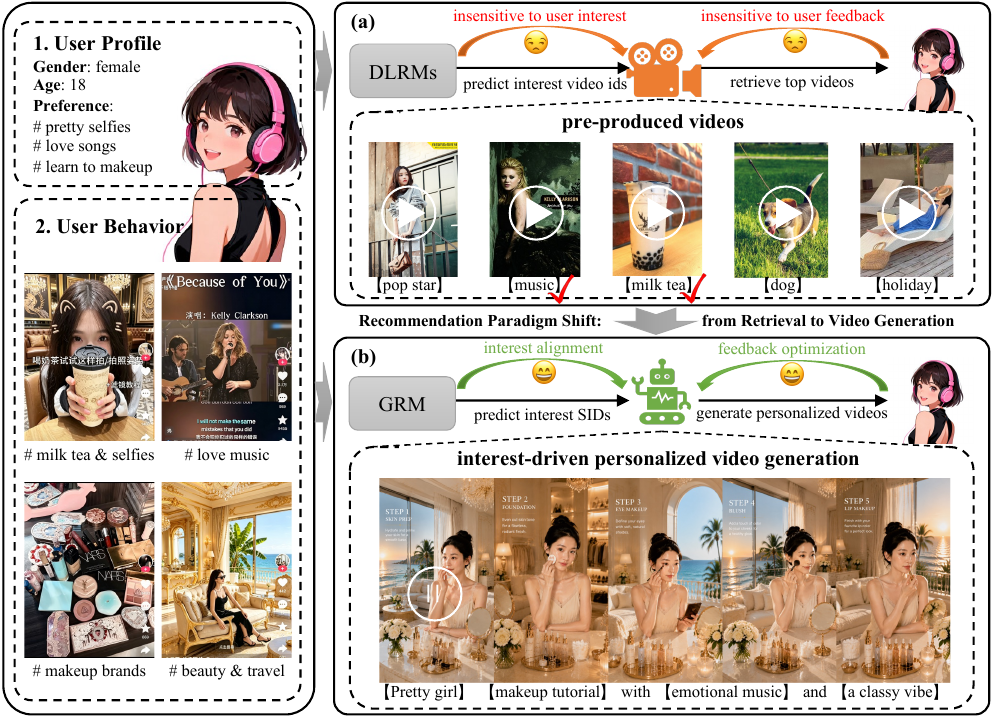}
        \caption{\small 
        Recommendation paradigm shift.
        (a) DLRMs retrieve videos from a fixed content pool, leading to suboptimal matches when user interests fall outside the pool;
        (b) Our paradigm generates personalized videos on demand that both align with the user interests predicted by a GRM and are driven by real user feedback in a closed loop, breaking the fixed-pool limit.
        }
    \label{fig:rag}
\end{figure}

Over the past decade, industrial video recommendation systems have followed a content-first paradigm, where videos are produced offline and recommendation models retrieve and rank items from a fixed pool. Deep learning recommendation models (DLRMs)~\cite{covington2016deep, cheng2016wide, zhou2018din, zhu2018learning} improve matching accuracy under this setting. More recently, generative recommendation models (GRMs)~\cite{deng2025onerec, xue2026gr4ad} extend this paradigm by modeling user interest through large-scale autoregressive generation over semantic IDs (SIDs)~\cite{rajput2023recommender}.

Despite these advances, existing systems remain fundamentally constrained by a static pool of pre-produced videos. Recommendation models can only retrieve the best available content from the existing pool, even when user interests fall outside the pool. This limitation is particularly severe in modern short-video platforms, where user interests are more dynamic, long-tailed, and diverse. As a result, retrieval-based systems are inherently limited in faithfully capturing fine-grained user interest.

Meanwhile, recent breakthroughs in AI-generated content (AIGC)\\~\cite{sora2_system_card, veo3_blog, team2025kling, wan2025, seedance2025seedance} have demonstrated unprecedented capabilities in open-domain video creation. Modern video generation models can produce cinematic-quality visual content with strong semantic controllability, opening up a new opportunity for recommendation:

\textit{Can recommendation systems move beyond retrieving existing videos to directly generate personalized videos from inferred user interests?}

Answering this question requires addressing two key challenges in recommendation and generation systems.

\textbf{The first challenge is how to bridge recommendation and generation into a unified framework.}
Recommendation models are trained on heterogeneous and discrete data, including user profiles, item features, and user behaviors, aiming to predict user interests. In contrast, video generation models operate on multimodal continuous signals, such as text, images, audio, and motion, focusing on generating coherent and high-fidelity videos. Given such fundamental differences in data representation and learning objectives, recommendation and generation are typically developed as two separate tasks, making it difficult to translate predicted user interests into controllable video generation. This separation also blocks user feedback from flowing back into the generation process, limiting the diversity and interest-alignment of the produced content.

\textbf{The second challenge is how to generate high-quality and interest-aligned videos at industrial scale.}
Although recent state-of-the-art video generation models~\cite{sora2_system_card, veo3_blog, team2025kling, wan2025, seedance2025seedance} achieve strong visual quality, they remain difficult to deploy in large-scale recommendation systems. These models often rely on manual prompting, multi-stage refinement and post-processing with professional tools, resulting in high latency and computational cost to produce a single user-satisfactory video. Personalizing across the diverse and long-tailed interests of hundreds of millions of users further amplifies these costs, making it infeasible to deploy such models directly in production.

To address these challenges, we propose \textbf{Recommendation-as-Generation} (\textbf{RaG}), a new paradigm that unifies recommendation and personalized video generation in a closed-loop framework, as illustrated in Figure~\ref{fig:rag}. Instead of retrieving from a fixed pool, RaG generates personalized videos directly from inferred user interests.

A key idea of RaG is to use \textbf{Disentangled Semantic IDs} (\textbf{D-SIDs}) as a unified interface between recommendation and generation. A multimodal large language model encodes each video into two factorized embeddings---one for \emph{content} (entities, topics) and the other for \emph{creative} attributes (style, rhythm, atmosphere). These embeddings are then quantized into discrete \emph{content SIDs} and \emph{creative SIDs}, jointly forming the video's D-SIDs. On the recommendation side, a generative recommendation  model (GRM) autoregressively predicts the D-SIDs of user interests. On the generation side, the predicted D-SIDs are decoded into personalized videos, connecting fine-grained interest modeling with controllable video generation.



To realize controllable video generation at scale, RaG develops \textbf{Video Generation Agents} (\textbf{VGAs}). Compared to monolithic, high-cost diffusion-based or prompt-engineering-heavy pipelines, VGAs adopt a hierarchical planning and refinement framework. Conditioned on user-interest D-SIDs, a fine-tuned \textbf{Instruction Model} (\textbf{IM}) first translates them into structured generation blueprints. Three role-specialized agents then reason and act over the blueprints, jointly modeling visual composition, audio alignment, and artistic effects. The three agents share a single LLM backbone and are jointly trained end-to-end, differentiated only through prompts and tool access. After the agents complete the pipeline, a bounded reflection loop (capped at two iterations) refines cross-modal consistency, balancing output quality with generation efficiency. The shared backbone further enables KV-cache reuse across agents to substantially accelerate inference. Combined with an SID-indexed cache that amortizes generation cost, VGAs reliably serve recommendation requests for hundreds of millions of users at industrial scale.

To close the optimization loop, RaG introduces \textbf{Synergistic Cross-Domain Reward Learning} (\textbf{SCRL}). Instead of naive reward aggregation that conflates heterogeneous reward signals, SCRL formulates multi-objective optimization as a constrained policy learning problem: user feedback serves as the primary objective, while interest alignment and video quality act as constraints. Group-decoupled reward normalization (GDPO~\cite{liu2026gdpo}) is applied per channel to reconcile scale mismatch, followed by a PID-controlled Lagrangian update~\cite{stooke2020responsive} to stabilize training. Together, SCRL unifies recommendation and video generation into a single closed-loop optimization where user interests, content quality, and real-world feedback co-evolve.

We deploy RaG on a large-scale production platform serving over 400 million daily active users in a revenue-critical advertising scenario. Online A/B testing shows significant improvements in ad revenue, validating the effectiveness of generation-driven personalization for recommendation. To the best of our knowledge, this is the first production-scale system that effectively unifies recommendation and personalized video generation. 

Our main contributions are summarized as follows:








\begin{itemize}[leftmargin=*,itemsep=1pt]
\item We propose \textbf{Recommendation-as-Generation} (\textbf{RaG}), a new paradigm that shifts recommendation from retrieving videos within a fixed pool to generating personalized videos directly from inferred user interests. Disentangled Semantic IDs (D-SIDs) serve as the unified latent interface between recommendation and generation, and Synergistic Cross-Domain Reward Learning (SCRL) closes the loop by enforcing interest alignment, user feedback, and video quality assessment.

\item We develop industrial-scale \textbf{Video Generation Agents} (\textbf{VGAs}) with hierarchical planning, collaborative multi-agent execution, and iterative refinement, enabling scalable and high-quality personalized video production.

\item Extensive offline experiments and online A/B testing on a production platform demonstrate substantial improvements in ad revenue, validating the effectiveness of large-scale personalized video generation for recommendation.

\end{itemize}

%% file: kdd/paradigm.tex
\begin{figure*}[t]
  \centering
  \includegraphics[width=1\textwidth]{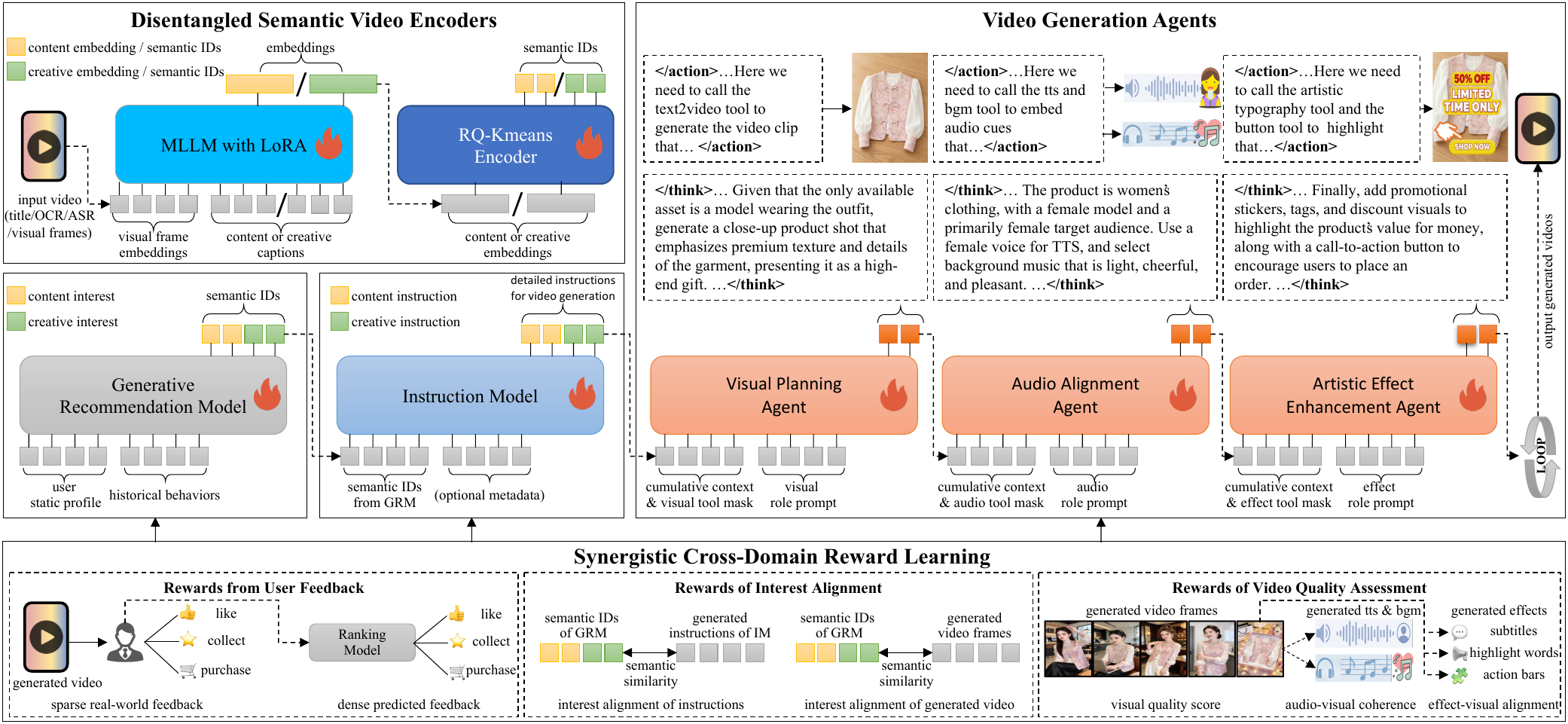}
  \caption {
  \small 
Architecture of the Recommendation-as-Generation (RaG) framework. Videos are encoded into Disentangled Semantic IDs (D-SIDs) that decouple content and creative semantics, forming a shared latent interface for recommendation and generation. The Generative Recommendation Model (GRM) predicts a user's interest D-SIDs from user context. The Instruction Model (IM) then translates these predicted D-SIDs, together with optional metadata, into shot-level production instructions, which are executed by the Video Generation Agents (VGAs) through hierarchical planning and refinement. The full pipeline is jointly optimized under Synergistic Cross-Domain Reward Learning (SCRL).
  }
  \label{fig:sys_pipeline}
\end{figure*}

\section{Methodology}

\subsection{Paradigm Shift: Recommendation as Generation}

Conventional recommendation systems~\cite{covington2016deep, cheng2016wide, zhou2018din, zhu2018learning} retrieve or rank videos from a fixed content pool.
Recent generative recommendation models (GRMs) formulate recommendation as autoregressive token prediction~\cite{xue2026gr4ad,deng2025onerec}, but still retrieve videos from the existing content pool set according to the predicted tokens.
As a result, these approaches remain limited by content coverage, often yielding suboptimal recommendations when user interests involve novel or long-tail semantics.

To overcome this limitation, we introduce the \textbf{Recommendation-as-Generation} (\textbf{RaG}) paradigm, which reformulates recommendation as an interest-conditioned video generation problem (Figure~\ref{fig:sys_pipeline}).
Instead of retrieving existing videos, RaG directly generates personalized videos from inferred user interests.
One key idea is to unify recommendation and video generation within a shared discrete latent space.

We construct this space using \textbf{Disentangled Semantic Video Encoders} (Section~\ref{sec:SID}), which map videos into disentangled semantic IDs (D-SIDs). These D-SIDs capture both semantic content and creative attributes, enabling both fine-grained interest modeling and controllable video generation.
Given a video $v$, the encoder $\mathcal{E}$ produces a sequence of tokens:
\begin{equation}
\mathbf{D{-}SIDs} = \mathcal{E}(v) = (s_\text{content}^1, \ldots, s_\text{content}^L, s_\text{creative}^1, \ldots, s_\text{creative}^L),
\end{equation}
which jointly represent video semantics and creative structure.

Building on this semantic space, recommendation is recast as generative interest modeling: given a user's profile and interaction history, the Generative Recommendation Model (GRM, Appendix~\ref{app:GRM}) autoregressively predicts the sequential D-SIDs representing the user's future interests:
\begin{equation}
\label{eq:rag_sid}
p(\mathbf{D{-}SIDs} \mid \mathbf{c}_{\text{user}})
=
\prod_{t=1}^{2L}
p(s_t \mid s_{<t}, \mathbf{c}_{\text{user}}),
\end{equation}
where $\mathbf{c}_{\text{user}}$ denotes the user context.

Unlike prior GRM-based approaches that use predicted D-SIDs as retrieval keys, we treat D-SIDs as \emph{generative interest representations} that can be directly decoded into new content, beyond a fixed pool. The overall pipeline is:
\begin{align}
\mathbf{D{-}SIDs} = \mathcal{E}(v)
\rightarrow
p(\mathbf{D{-}SIDs} \mid \mathbf{c}_{\text{user}})
\rightarrow
\hat{v} = \mathcal{G}(\mathbf{D{-}SIDs}),
\end{align}
where user interests are modeled in the latent semantic space and decoded into personalized videos.
However, directly optimizing $\mathcal{G}$ for both generation quality and interest alignment is challenging.
We therefore decompose the generation process into a hierarchical framework.

We introduce an \textbf{Instruction Model} (Section~\ref{sec:IM}) that translates D-SIDs into natural language instructions, providing interpretable and structured guidance for downstream agents. Building on this, we develop \textbf{Video Generation Agents} (Section~\ref{sec:A-VGS}) that generate videos through collaborative agents, enabling hierarchical planning, multimodal alignment, artistic enhancement, and iterative refinement. Finally, we optimize the entire framework via \textbf{Synergistic Cross-Domain Reward Learning} (Section~\ref{sec:Reward}), jointly capturing user interest alignment, generation quality, and user engagement signals.

%% file: kdd/3Method.tex
\subsection{Disentangled Semantic Video Encoders}\label{sec:SID}
\subsubsection{Multimodal Representation Learning}
Building upon Qwen2.5-VL-7B-Instruct~\cite{qwen2_5_vl}, we propose an instruction-guided disentangled representation framework that separates semantic content and creative attributes from the same video. For multimodal input processing, we directly reuse Qwen2.5-VL's native visual encoder and text tokenizer.

We first extract its visual token representations using the vision encoder:
$H = \mathcal{F}(v),H \in \mathbb{R}^{N \times d},$
where $H$ denotes a sequence of visual tokens capturing spatial-temporal semantics.

To obtain disentangled signals, we leverage our in-house dense captioning model (CapModel) to generate factor-specific textual descriptions:
\begin{equation}
D_m = \text{CapModel}(v,\text{PROMPT}_m),
\quad
m \in \{\text{content}, \text{creative}\},
\end{equation}
where $D_\text{content}$ describes semantic content (entities, topics), while $D_\text{creative}$ captures creative attributes (style, rhythm and atmosphere).

The instructions are encoded via the text tokenizer:
$Q_m = \mathcal{T}(D_m)$, $Q_m \in \mathbb{R}^{L_m \times d}$, where $L_m$ is the instruction length.
We obtain multimodal representations by jointly encoding visual and textual inputs with Qwen2.5-VL-7B-Instruct, and use the last-token hidden state of the final layer as the pooled multimodal representation:
\begin{equation}
\mathbf{z}_m
=
\mathrm{Normalize}(\mathrm{VLM}(H, Q_m)),
\quad
\mathbf{z}_m \in \mathbb{R}^{d},\ \|\mathbf{z}_m\|_2 = 1,
\end{equation}
yielding $\mathbf{z}_\text{content}$ and $\mathbf{z}_\text{creative}$ as L2-normalized content and creative representations, respectively.


To encourage representation consistency, we employ a contrastive loss for each module:
\begin{equation}
\mathcal{L}_m
=
-\log
\frac{
\exp(\mathrm{sim}(\mathbf{z}_m^i,\mathbf{z}_m^j)/\tau)
}{
\sum_k \exp(\mathrm{sim}(\mathbf{z}_m^i,\mathbf{z}_m^k)/\tau)
},
\end{equation}
where $\mathbf{z}_m^j$ is the positive pair of $\mathbf{z}_m^i$ within a batch, and k indexes all candidates including the positive.

To reduce cross-factor leakage, we introduce an orthogonality constraint:
\begin{equation}
\mathcal{L}_\text{orth}
=
\|\mathbf{z}_\text{content}^\top \mathbf{z}_\text{creative}\|_2^2.
\end{equation}

The final objective is:
\begin{equation}
\mathcal{L}
=
\mathcal{L}_\text{content}
+
{\gamma_1} \mathcal{L}_\text{creative}
+
{\gamma_2} \mathcal{L}_\text{orth}.
\end{equation}

\subsubsection{Discrete Tokenization}

To facilitate generative recommendation within the latent space, we discretize the disentangled multimodal representations into semantic IDs.

Specifically, each representation $\mathbf{z}_m$ is independently quantized via Residual Quantization (RQ)-based K-means~\cite{luo2024qarm}, yielding a quantized embedding $\mathbf{e}_m$ that approximates $\mathbf{z}_m$ as a sum of codebook vectors across $L$ hierarchical layers:
\begin{equation}
\mathbf{e}_m = \sum_{l=1}^{L} \mathbf{c}_{m}^{l}(s_{m}^{l}) \,\approx\, \mathbf{z}_m,
\qquad
\mathbf{e}_m \in \mathbb{R}^{d},
\label{eq:rq_quantization}
\end{equation}
where $s_{m}^{l}$ denotes the discrete code index at layer $l$ for modality $m$, and $\mathbf{c}_{m}^{l}(\cdot)$ is the corresponding codebook lookup. Each modality maintains an independent codebook with 8,192 entries per layer. The final D-SIDs are obtained by concatenating the per-modality code sequences: $\mathbf{D{\text{-}}SIDs} = \bigl[\,s_{\text{content}}^{\,1:L}\,;\, s_{\text{creative}}^{\,1:L}\,\bigr]$.

\subsection{Instruction Model}\label{sec:IM}
The Instruction Model translates disentangled semantic IDs into shot-level video production instructions. Unlike conventional caption generation, these instructions explicitly specify scene composition, camera motion, temporal pacing, and cinematic style, serving as an intermediate semantic bridge between discrete user interests and controllable video generation.

\subsubsection{Supervision Construction.}
Since no off-the-shelf dataset contains video-instruction pairs at the shot level, we distill supervision from a strong multimodal teacher. For each video $v$, we first extract its D-SIDs (Section~\ref{sec:SID}), and then prompt Gemini2.5~Pro~\cite{comanici2025gemini} with a carefully designed instruction template $\text{PROMPT}_{\text{inst}}$ to produce the target shot-level script:
\begin{equation}
D_{\text{inst}}
\;=\;
\text{Gemini}(v,\,\text{PROMPT}_{\text{inst}})
\;=\;
(y_{1},\, y_{2},\, \ldots,\, y_{L_{\text{inst}}}),
\end{equation}
where $D_{\text{inst}}$ is a token sequence of length $L_{\text{inst}}$ serving as ground-truth supervision. To accommodate advertising scenarios where the generated video must reflect specific products being promoted, we further introduce an \emph{optional} metadata factor $D_{\text{meta}}$ (e.g., product information and marketing topics) as an auxiliary conditioning signal. When unavailable (e.g., for pure organic videos), $D_{\text{meta}}$ is simply masked, leaving instruction generation conditioned on D-SIDs alone.

\subsubsection{Model and Optimization Objective.}
We instantiate the Instruction Model with Qwen3-8B~\cite{yang2025qwen3}, which consumes two heterogeneous token sequences---the primary D-SIDs and the auxiliary $D_{\text{meta}}$ (masked when unavailable)---mapped into the LLM's input embedding space and concatenated as the prefix.

For D-SIDs, we reconstruct continuous embeddings from the discrete codes via reverse RQ-Kmeans, $\mathbf{e}_{\text{D-SIDs}} = [\mathbf{e}_{\text{content}};\, \mathbf{e}_{\text{creative}}] \in \mathbb{R}^{2 \times d}$, and map them through a learnable projector $\phi(\cdot)$ to $\mathbf{h}_{\text{D-SIDs}} = \phi(\mathbf{e}_{\text{D-SIDs}}) \in \mathbb{R}^{2 \times d'}$. For metadata, $D_{\text{meta}}$ is tokenized and embedded by the LLM's native text tokenizer $\mathcal{T}$ to $Q_{\text{meta}} = \mathcal{T}(D_{\text{meta}}) \in \mathbb{R}^{L_{\text{meta}} \times d'}$, with $L_{\text{meta}}$ denoting the token length. Conditioned on both, the model autoregressively predicts the instruction sequence
\begin{equation}
\hat{D}_{\text{inst}} \;=\; \text{LLM}(\mathbf{h}_{\text{D-SIDs}},\, Q_{\text{meta}}),
\end{equation}
and is optimized with a standard next-token prediction loss against the Gemini-distilled supervision:
\begin{equation}
\mathcal{L}_{\text{NTP}}
\;=\;
- \sum_{t=1}^{L_{\text{inst}}}
\log P\!\bigl(y_{t}\,\big|\,y_{<t},\, \mathbf{h}_{\text{D-SIDs}},\, Q_{\text{meta}}\bigr),
\end{equation}
so that $\hat{D}_{\text{inst}}$ token-wise approximates $D_{\text{inst}}$.

\subsubsection{Three-Stage Training.}
We adopt a three-stage training strategy: in the first stage, the backbone LLM is frozen and only the projector $\phi(\cdot)$ is optimized to align D-SIDs' embeddings with the language space; in the second stage, both the projector and LLM parameters are jointly fine-tuned for improved semantic fidelity and controllable instruction generation; in the third stage, we further enhance the model using reinforcement learning with reward optimization, as described in Section~\ref{sec:Reward}.

\subsection{Video Generation Agents}\label{sec:A-VGS}
As illustrated in Figure~\ref{fig:sys_pipeline}, industrial-scale personalized video generation cannot be effectively handled by a monolithic generator. The one-shot production of visuals, audio, and effects often leads to semantic inconsistency and limited controllability. Moreover, video production exhibits a hierarchical dependency structure where visual planning determines the narrative flow, while audio and effects are conditioned on the visual state.

To address this, we propose Video Generation Agents (VGAs), formulated as a structured multi-agent decision process over an evolving generation state.

\vspace{2pt}
\noindent\textbf{Agentic Formulation.}
We model video generation as a sequential decision process executed by a team of sub-agents. At each step $t$, the active sub-agent observes a state $\mathcal{S}_t$, selects an action $a_t$ according to its policy $\pi_{\theta}$, and transitions to the next state via a deterministic operator $\mathcal{P}$:
\begin{equation}
a_t \sim \pi_{\theta}(a_t \mid \mathcal{S}_t),
\qquad
\mathcal{S}_{t+1} = \mathcal{P}(\mathcal{S}_t, a_t).
\end{equation}
To enable efficient backbone reuse across sub-agents (detailed later), we serialize the state as an ordered prefix followed by stage-dependent tokens:
\begin{equation}
\mathcal{S}_t = \bigl[\,\underbrace{\hat{D}_{\text{inst}};\, D_{\text{tool}}}_{\text{shared prefix}};\; \underbrace{\mathcal{O}_{<t};\, \text{PROMPT}_\text{role}}_{\text{stage-dependent}}\,\bigr],
\label{eq:vga_state}
\end{equation}
where $\hat{D}_{\text{inst}}$ is the instruction sequence produced by the Instruction Model; $D_{\text{tool}}$ is the description of all available tools, including in-house pretrained text-to-video and image-to-video models and external audio synthesis and visual effect APIs; $\mathcal{O}_{<t}$ is the running concatenation of all earlier sub-agents' role prompts and their generated outputs; and $\text{PROMPT}_\text{role}$ is a short role-specific prompt that activates the current sub-agent. The action $a_t$ corresponds to a modality-specific intent for visual, audio, or effect generation.

VGAs consist of three role-specialized sub-agents, each acting according to its own policy:
\begin{equation}
\pi_{\theta}(a_t \mid \mathcal{S}_t) = \{\pi_{\text{visual}},\, \pi_{\text{audio}},\, \pi_{\text{effect}}\}(a_t \mid \mathcal{S}_t),
\end{equation}
corresponding to visual planning, audio synthesis, and post-production effects, respectively. We next describe each sub-agent in turn.

\vspace{2pt}
\noindent\textbf{1. Visual Planning Agent (VPA).}
At the visual stage, $\text{PROMPT}_\text{role}$= $\text{PROMPT}_\text{visual}$ and $\mathcal{O}_{<t}$ is empty (or carries the previous reflection round's content). The VPA acts as the global controller, producing a clip-level storyboard with scene segments, layout configurations, and temporal boundaries: $\mathcal{I}_\text{visual} = \pi_{\text{visual}}(\mathcal{S}_t)$.

\vspace{2pt}
\noindent\textbf{2. Audio Alignment Agent (AAA).}
At the audio stage, $\text{PROMPT}_\text{role}$ = $\text{PROMPT}_\text{audio}$ and $\mathcal{O}_{<t}$ extends with $(\text{PROMPT}_\text{visual}, \mathcal{I}_\text{visual})$. The AAA generates temporally aligned audio (speech and music) synchronized with scene transitions: $\mathcal{I}_\text{audio} = \pi_{\text{audio}}(\mathcal{S}_t)$.

\vspace{2pt}
\noindent\textbf{3. Artistic Effect Enhancement Agent (AEEA).}
At the effect stage, $\text{PROMPT}_\text{role} = \text{PROMPT}_\text{effect}$ and $\mathcal{O}_{<t}$ further extends with $(\text{PROMPT}_\text{audio}, \mathcal{I}_\text{audio})$. The AEEA performs post-production refinement by adding subtitles, visual effects, transitions, and call-to-action elements: $\mathcal{I}_\text{effect} = \pi_{\text{effect}}(\mathcal{S}_t)$.

\vspace{2pt}
\noindent\textbf{Hierarchical Generation with Bounded Reflection.}
The three intent outputs are composed into the final video via a unified generation operator $\mathcal{G}$:
\begin{equation}
\mathcal{V} = \mathcal{G}(\mathcal{I}_\text{visual},\, \mathcal{I}_\text{audio},\, \mathcal{I}_\text{effect}).
\end{equation}
To improve cross-modal consistency, VGAs operate within a bounded reflection loop that follows the standard Observe$\rightarrow$Think$\rightarrow$Act cycle, capped at two iterations to balance output quality with generation efficiency.

\vspace{2pt}
\noindent\textbf{Shared Backbone and KV-Cache Reuse.}
Although VGAs comprise three role-specific sub-agents, they are not three separate models---all share a single Qwen2.5-32B backbone~\cite{yang2024qwen25} with fully shared parameters. Differentiation arises purely from the state $\mathcal{S}_t$ in Eq.~\eqref{eq:vga_state}: $\text{PROMPT}_{\text{role}}$ activates the target sub-agent, an attention mask over $D_{\text{tool}}$ restricts it to the accessible tool subset, and $\mathcal{O}_{<t}$ supplies the cumulative upstream context. This serialization also enables straightforward KV-cache reuse: with sub-agents invoked sequentially and $\mathcal{O}_{<t}$ growing append-only, every previously generated token stays in the KV cache, leaving each downstream sub-agent to only encode its own $\text{PROMPT}_{\text{role}}$, substantially reducing per-request inference latency.

The agent policies are optimized via synergistic cross-domain reward signals that jointly capture generation quality, interest alignment, and user feedback, as detailed in Section~\ref{sec:Reward}.

\subsection{Synergistic Cross-Domain Reward Learning}\label{sec:Reward}
\subsubsection{Cross-Domain Reward Formulation}
To further enhance both recommendation and video generation performance, we formulate a structured cross-domain reward scheme with three synergistic objectives: \textit{video quality}, \textit{interest alignment}, and \textit{user feedback}. Unless otherwise specified, all reward models share the same Transformer-based architecture and are trained on task-specific datasets.

\vspace{2pt}
\noindent\textbf{1. Video Quality Rewards.}

We evaluate the perceptual and compositional quality of generated videos from three complementary aspects: visual quality, audio-visual coherence, and effect-visual alignment. Formally, we define:
$R_{\text{quality}} = R_{\text{visual}} + R_{\text{audio}} + R_{\text{effect}}$, 
where:
\begin{itemize}[leftmargin=*]
    \item $R_{\text{visual}}$ evaluates visual quality, including aesthetic appeal and spatio-temporal consistency to ensure coherent motion and stable rendering,
    \item $R_{\text{audio}}$ measures alignment between audio and visual content, covering both speech synchronization (TTS) and background music consistency (BGM),
    \item $R_{\text{effect}}$ captures the quality and alignment of visual effects, including subtitles, highlights, and interactive elements such as action bars.
\end{itemize}

\vspace{2pt}
\noindent\textbf{2. Interest Alignment Rewards.}

To keep generated content aligned with user interests throughout the pipeline, we apply alignment rewards at multiple stages, anchored on the D-SIDs that encode user interests in a structured latent space:
$R_{\text{align}} = R_{\text{instr-align}} + R_{\text{rep-align}}$, where: 
\begin{itemize}[leftmargin=*]
    \item $R_{\text{instr-align}}$ enforces semantic consistency between GRM-generated D-SIDs and the generated instructions,
    \item $R_{\text{rep-align}}$ measures semantic similarity between GRM-generated D-SIDs and the generated videos.
\end{itemize}

\vspace{2pt}
\noindent\textbf{3. User Feedback Rewards.}

To enhance downstream user engagement, we leverage user interaction signals such as clicks and conversions as the core reward for optimization. However, real-world interaction signals are sparse and delayed, making them insufficient for stable and efficient policy optimization.

To mitigate this issue, we augment sparse interaction signals with dense engagement estimates from deployed ranking models. We define the overall reward as:
$R_{\text{feedback}} = R_{\text{real}} + R_{\text{pred}}$,
where:
\begin{itemize}[leftmargin=*]
    \item $R_{\text{real}}$ denotes sparse but high-fidelity user interaction signals observed from real feedback, including behaviors such as click, like, collect, and purchase,
    \item $R_{\text{pred}}$ denotes dense engagement signals estimated by ranking models, which capture user preference strength beyond explicit interactions.
\end{itemize}

\subsubsection{Constrained Policy Optimization with GDPO}
\label{sec:cpo-gdpo}
To jointly optimize the heterogeneous, cross-domain rewards introduced above, we formulate reward learning within the RaG framework as a \emph{constrained policy optimization} problem solved by GDPO~\cite{liu2026gdpo}. The design addresses two practical challenges in multi-reward RL: (i)~the scale mismatch and optimization instability caused by heterogeneous reward signals, and (ii)~the difficulty of statically balancing competing objectives without sacrificing the dominant goal.

\paragraph{Problem setup.}
Given an input context $x$, the policy $\pi_\theta$ samples a candidate set $\mathcal{Y} = \{y_1, \dots, y_K\} \sim \pi_\theta(\cdot \mid x)$. Each candidate $y_i$ is evaluated by a collection of heterogeneous reward functions covering user feedback $R_{\text{feedback}}(y_i)$, interest alignment $R_{\text{align}}(y_i)$, and video quality $R_{\text{quality}}(y_i)$. These rewards differ in scale, density, and reliability, motivating the constrained formulation below.

\paragraph{Constrained reward formulation.}
We designate user feedback as the primary objective and treat interest alignment and video quality as inequality constraints with target thresholds $\tau_{a(lign)}$ and $\tau_{q(uality)}$. The composite reward for each candidate $y_i$ is defined as
\begin{equation}
\label{eq:constrained-reward}
R(y_i) = R_{\text{feedback}}(y_i)
- \!\!\!\sum_{c \in \{a, q\}}\!\!\! \lambda_c(t)\,\mathrm{ReLU}\!\big(\tau_c - R_c(y_i)\big),
\end{equation}
where $\lambda_a(t),\lambda_q(t)\ge 0$ are time-varying Lagrangian multipliers, updated via a PID-controlled rule on constraint violations~\cite{stooke2020responsive} to avoid the oscillation and overshoot of naive primal--dual updates. To avoid hand-tuned magic numbers, we calibrate each threshold relative to the SFT baseline distribution on a held-out validation set as $\tau_c = \mu_c^{\text{base}} + k_c\,\sigma_c^{\text{base}}$, where the strictness factor $k_c$ encodes the module's role in RaG: VGAs adopt the strictest setting ($k_c=1.1$ for both $\tau_a$ and $\tau_q$) as it directly governs final video generation; IM retains a comparable $\tau_a$ ($k_a=0.8$) to enforce instruction-level alignment; while GRM applies a relaxed $\tau_a$ ($k_a=0.3$), with the video-quality constraint omitted for the latter two modules.


\paragraph{Group-decoupled normalization and advantage.}
Given the constrained reward in Eq.~\eqref{eq:constrained-reward}, GDPO further eliminates residual scale mismatch among reward channels via per-reward standardization prior to aggregation, and computes a group-relative advantage over the sampled candidate set $\mathcal{Y}$:
\begin{equation}
\label{eq:group-adv}
A_i = \frac{R(y_i) - \mu(\mathcal{Y})}{\sigma(\mathcal{Y}) + \epsilon},
\end{equation}
where $\mu(\mathcal{Y})$ and $\sigma(\mathcal{Y})$ denote the group-level mean and standard deviation of the rewards over $\mathcal{Y}$. This decoupled normalization stabilizes optimization across rewards with disparate magnitudes.

\paragraph{Optimization objective.}
The policy is updated by maximizing the group-relative advantage anchored to the frozen SFT policy $\pi_{\text{ref}}$:
\begin{equation}
\label{eq:gdpo-loss}
\mathcal{L}_{\text{GDPO}} =
- \mathbb{E}_{(x, y_i)}
\left[
A_i \,\log \frac{\pi_\theta(y_i \mid x)}{\pi_{\text{ref}}(y_i \mid x)}
\right].
\end{equation}
For brevity, we omit the importance-sampling ratio clipping and the KL regularization term against $\pi_{\text{ref}}$ that are commonly used to stabilize policy optimization; both are retained in our implementation and follow the standard GDPO formulation~\cite{liu2026gdpo}.

%% file: kdd/Deployment.tex
\section{Deployment}\label{sec:Deployment}

\begin{figure}[t]
        \centering
        \includegraphics[width=0.48\textwidth]{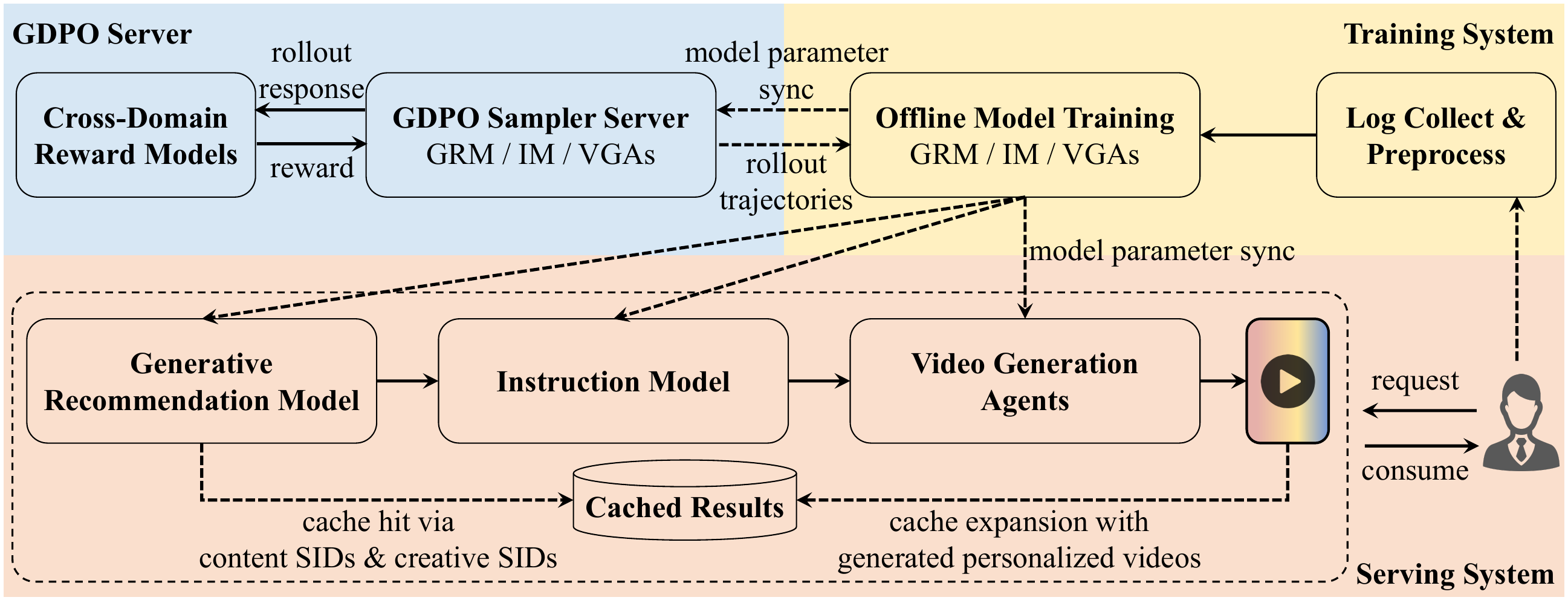}
    \caption{\small Training and serving architecture of the proposed Recommendation-as-Generation system.}
    \label{fig:deployment}
\end{figure}

We deploy RaG in Kuaishou’s large-scale advertising system, serving over 400 million users (Figure~\ref{fig:deployment}). The system unifies real-time user interest modeling with large-scale personalized video generation under strict latency constraints. Since video generation is orders of magnitude slower than interest inference (Appendix~\ref{app:runtime}), we design a decoupled deployment architecture to bridge this efficiency gap while maintaining end-to-end personalization quality. The system consists of three decoupled modules: real-time interest modeling, nearline video generation, and latency-aware serving.



\vspace{2pt}
\noindent\textbf{Real-Time Interest Modeling.}
The Generative Recommendation Model (GRM) is continuously trained on streaming user interaction logs (impression, click, watch time, and conversion) to adapt to non-stationary user behavior, combining streaming supervised updates with periodic GDPO-based optimization.

At real-time inference, GRM performs low-latency autoregressive generation of structured Semantic IDs (SIDs), which encode user interests as semantic targets for downstream content generation.

\vspace{2pt}
\noindent\textbf{Nearline Video Generation.}
The Instruction Model (IM) and Video Generation Agents (VGAs) are trained on large-scale agentic supervision data curated from high-quality videos, and optimized via supervised fine-tuning followed by constrained GDPO to jointly improve generation quality and interest alignment. Both models are periodically updated in full-batch mode to ensure training stability while adapting to evolving user interests and emerging video patterns.

At serving time, conditioned on GRM-generated SIDs, IM and VGAs operate in a nearline pipeline to generate personalized videos. To handle the heavy generation load, VGAs apply \emph{KV-cache reuse} as established in Section~\ref{sec:A-VGS}: with sub-agents invoked sequentially over an append-only state, every previously generated token stays in the KV cache, leaving only each sub-agent's own short $\text{PROMPT}_{\text{role}}$ to be encoded per call, substantially reducing per-request inference latency. The outputs are continuously accumulated into a growing personalized video space, enabling coverage expansion while decoupling video generation from real-time serving.

\vspace{2pt}
\noindent\textbf{Latency-Aware Serving.}
To meet real-time consumption requirements in recommendation scenarios, the system adopts a hierarchical serving strategy organized around whether the requested content-level SIDs are covered by the cache.

\emph{Case 1: content-SIDs hit.} If the matched cache entry also covers the creative-level SIDs, the system returns the previously generated video directly with negligible latency; otherwise, it serves a content-consistent cached video while asynchronously scheduling the missing creative variations, with higher-frequency creatives prioritized in the generation queue.

\emph{Case 2: content-SIDs miss.} The system serves videos associated with the nearest-neighbor SIDs for immediate consumption, while enqueuing the uncovered SIDs for prioritized future generation.

%% file: kdd/4Experiment.tex
\begin{table}[t]
\caption{\small Online A/B test results. Rev. denotes ad revenue. Results are reported as relative improvements over production baselines.}
\label{tab:ab_style3}
\resizebox{1.0\columnwidth}{!}{%
\centering
\footnotesize

\begin{tabular}{l@{\hspace{6pt}}c@{\hspace{6pt}}c}
\toprule
\textbf{Method} 
& \begin{tabular}[c]{@{}c@{}}\textbf{Rev. (\%$\uparrow$)}\\ {vs. DLRM baseline}\end{tabular}
& \begin{tabular}[c]{@{}c@{}}\textbf{Rev. (\%$\uparrow$)}\\ {vs. GRM baseline}\end{tabular}
\\
\midrule

\textbf{Production Baseline} \\
DLRM baseline & -- & -- \\
GRM baseline~\cite{xue2026gr4ad}  & +3.526\% & -- \\
\midrule

\textbf{Enhanced GRM} \\
GRM + Disentangled-SIDs (D-SIDs) & +4.460\% & +0.902\% \\
\midrule

\textbf{Full System (RaG)} \\
RaG (GRM + D-SIDs + IM + VGAs + SCRL)  & \textbf{+5.462\%} &
\textbf{+1.870\%}
\\

\bottomrule
\end{tabular}

}
\end{table}

\section{Experiments}
\subsection{Online A/B Testing}
We deploy the proposed Recommendation-as-Generation (RaG) framework in the real-world advertising platform of Kuaishou and conduct large-scale online A/B experiments to evaluate its industrial effectiveness. The experiments mainly focus on two aspects: 
\textbf{(1)} the effectiveness of Disentangled Semantic IDs (D-SIDs) for generative recommendation, and 
\textbf{(2)} the additional gains brought by SID-driven personalized video generation.


Table~\ref{tab:ab_style3} summarizes the online results. Replacing the production DLRM-based pipeline with the Generative Recommendation Model (GRM) yields consistent ad revenue gains, and the proposed Disentangled Semantic IDs (D-SIDs) further lift the improvement from +3.526\% to +4.460\%, confirming that decoupling content and creative semantics yields a more structured latent space and mitigates interference during autoregressive generation. Nevertheless, both variants remain within the retrieval paradigm, selecting candidates from a fixed pool.

\begin{table}[t]
\centering
\caption{\small Quality of the Disentangled SIDs. We report both (i) embedding-based semantic retrieval quality and (ii) SID discretization quality. Improvements over the strongest baseline are highlighted. \textit{Impr.}: improvement; \textit{R@k}: Recall@k; \textit{Cpr.}: compression rate; \textit{Col.}: collision rate.}
\label{tab:UA-DBS-Full}
\resizebox{1.0\columnwidth}{!}{%
\begin{tabular}{lcccc}
\toprule

\textbf{Method} 
& \multicolumn{3}{c}{\textbf{Semantic Retrieval (R@K)}} 
& \textbf{Discretization Quality} \\
\cmidrule(lr){2-4} \cmidrule(lr){5-5}

& \textbf{\textit{R@1} $\uparrow$} & \textbf{\textit{R@5} $\uparrow$} & \textbf{\textit{R@10} $\uparrow$} 
& \textbf{\textit{Cpr.} $\downarrow$ / \textit{Col.} $\downarrow$} \\
\midrule

VLM2Vec-V2~\cite{meng2025vlm2vec} 
& 0.485 & 0.690 & 0.756 
& -- \\

QARM~\cite{luo2024qarm} 
& 0.541 & 0.812 & 0.893 
& \underline{1.14} / \underline{18.24\%} \\

Qwen2.5-VL-7B~\cite{qwen3embedding} 
& \underline{0.769} & \underline{0.948} & \underline{0.977} 
& -- \\

\midrule

\textbf{Ours (D-SIDs)} 
& \textbf{0.896} 
& \textbf{0.985} 
& \textbf{0.994} 
& \textbf{1.02 / 2.62\%} \\

\midrule

\textbf{\textit{Impr.}} 
& \textbf{+16.5\%} 
& \textbf{+3.9\%} 
& \textbf{+1.7\%} 
& \textbf{-10.5\% / -15.6pp} \\

\bottomrule
\end{tabular}
}
\end{table}


Finally, the full RaG framework---integrating GRM, D-SIDs, the Instruction Model (IM), and Video Generation Agents (VGAs) under Synergistic Cross-Domain Reward Learning (SCRL)---delivers a +5.462\% ad revenue gain over the DLRM-based pipeline. Crucially, RaG also outperforms the strong GRM baseline by +1.870\%, with this additional lift coming directly from D-SIDs-driven personalized video generation. This marks a paradigm shift from retrieval-based to generation-based recommendation, where user interests actively drive personalized content production rather than merely matching existing candidates.

\subsection{Offline Ablation Studies}
We ablate the key components of RaG framework---D-SIDs, IM, VGAs, and SCRL optimization---to assess their contributions in terms of semantic representation quality, instruction generation capability, and reward-driven video generation performance.

\begin{table}[bt]
\centering
\caption{\small Evaluation of videos between the proposed VGAs vs. the workflow baseline. For Automated Score, we present average and median score.}
\small
\label{tab:consistency_success}
\resizebox{1.0\columnwidth}{!}{%
\begin{tabular}{lccc}
\toprule
\textbf{Metric} & \textbf{Workflow Baseline} & \textbf{VGAs} & \textbf{\textit{Impr.}}\\
\midrule
{Automated Score} $\uparrow$ & 62.4 / 62.0 & \textbf{71.3 / 76.0} & \textbf{+14.3\% / +22.6\%} \\
{Automated Win Rate} $\uparrow$ & 28.7\% & \textbf{70.1\%} & \textbf{+41.4pp} \\
{User Study Win Rate} $\uparrow$ & 34.4\% & \textbf{52.9\%} & \textbf{+18.5pp} \\
\bottomrule
\end{tabular}
}
\end{table}

\subsubsection{Quality of Disentangled SIDs} The D-SIDs consist of two core components, i.e., multimodal representation learning and semantic quantization; we systematically analyze their effectiveness in the following experiments.

\vspace{2pt}
\noindent\textbf{Multimodal Representation Learning.} We evaluate the proposed instruction-guided representation learning under a product-level retrieval setting to ensure fair comparison of semantic alignment capability. As shown in Table~\ref{tab:UA-DBS-Full}, our method consistently outperforms all baselines, achieving $0.896 / 0.985 / 0.994$ in R@1/5/10. In particular, R@1 improves by $+16.5\%$ over the strongest baseline (Qwen2.5-VL-7B), demonstrating stronger semantic discriminability under identical retrieval conditions.

\vspace{2pt}
\noindent\textbf{Semantic Quantization.} We construct the D-SIDs by applying RQ-KMeans residual quantization separately to the content and creative embeddings, yielding disentangled content SIDs and creative SIDs. For a fair comparison, both D-SIDs and QARM adopt an identical quantization setup with a 4-layer codebook and 8,192 codes per layer. As shown in Table~\ref{tab:UA-DBS-Full}, our method achieves superior discretization quality, reducing compression distortion to $1.02$ and collision rate to $2.62\%$. Compared to QARM ($1.14 / 18.24\%$), this corresponds to a $10.5\%$ reduction in compression error and a $15.6$pp lower collision rate, indicating a more compact and collision-resistant semantic space.






\begin{table}[bt]
\centering
\caption{\small Reward ablation with corresponding evaluation metrics. For each reward component, we report its dedicated metric on a corresponding evaluation set, comparing the policy trained with that reward (\textit{Ours}) against the no-reward base policy (\textit{Base}).}
\label{tab:reward_abs}
\resizebox{1.0\columnwidth}{!}{%
\small
\setlength{\tabcolsep}{4.5pt}
\renewcommand{\arraystretch}{1.1}

\begin{tabular}{ccc c}
\toprule

\multicolumn{3}{c}{\textbf{Video Quality Rewards}} 
& \textbf{Automated Win Rate $\uparrow$} \\
\midrule

$R_{\text{visual}}$ & $R_{\text{audio}}$ & $R_{\text{effect}}$ 
& Base $\rightarrow$ Ours \\
\cmidrule(lr){1-3}\cmidrule(lr){4-4}

\checkmark &  &  & 29.3\% $\rightarrow$ \textbf{50.7\% (+21.4pp)} \\
 & \checkmark &  & 24.0\% $\rightarrow$ \textbf{48.0\% (+24.0pp)} \\
 &  & \checkmark & 22.7\% $\rightarrow$ \textbf{41.3\% (+18.6pp)} \\
\checkmark & \checkmark & \checkmark & 37.3\% $\rightarrow$ \textbf{56.0\% (+18.7pp)} \\
\midrule

\multicolumn{3}{c}{\textbf{+ Interest Alignment Rewards}} 
& \textbf{Interest Alignment Score $\uparrow$} \\
\midrule

 & $R_{\text{align}}$ &  & Base $\rightarrow$ Ours \\
\cmidrule(lr){1-3}\cmidrule(lr){4-4}

 & \checkmark &  & 0.707 $\rightarrow$ \textbf{0.828 (+17.1\%)} \\



\bottomrule

\end{tabular}
}
\end{table}

\begin{figure*}[ht]
        \centering
        \includegraphics[width=\textwidth]{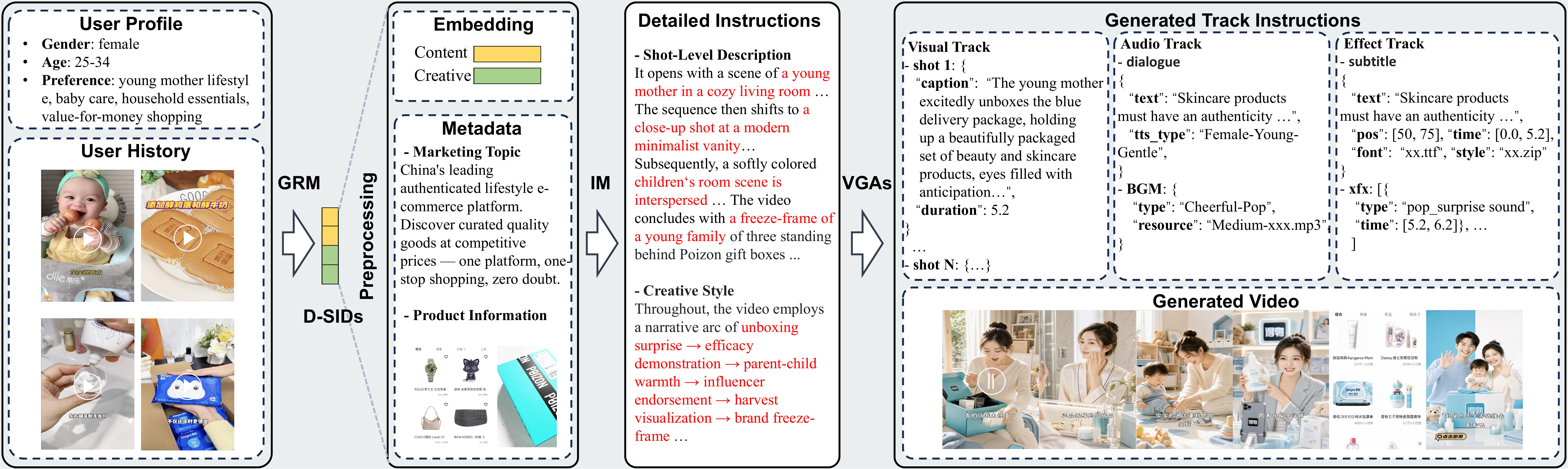}
    \caption{\small Qualitative example of interest-driven personalized video generation in advertising scenarios.}
    \label{fig:e2e_overview}
\end{figure*}

\subsubsection{Instruction Model Configuration}
We evaluate the Instruction Model in terms of decoding fidelity, measured by the cosine similarity between generated instructions and ground-truth summaries using Qwen3-Embedding-8B. Empirically, we observe that both increased training data and model capacity lead to consistent improvements. Specifically, the performance improves from $0.7760$ (8B model, 100K samples) to $0.8096$ (8B model, 1M samples), and further to $0.8212$ with a 32B model trained on 1M samples.

Considering the trade-off between performance and computational efficiency, we adopt the 8B model trained on 1M samples as the default configuration, which achieves competitive decoding fidelity while offering significantly lower deployment cost compared to larger models.

\subsubsection{Performance Analysis of Video Generation}

We evaluate the effectiveness of the proposed Video Generation Agents (VGAs) from two perspectives: (i) system-level comparison against conventional production pipelines, and (ii) the contribution of different reward components to optimization. Specifically, we assess three aspects: generation quality, including automated and human preference evaluations; and interest alignment score, measuring consistency between generated videos and target user interests. See Appendix~\ref{app:evaluation} for details.

\vspace{2pt}
\noindent\textbf{System-level Comparison.}
We compare VGAs with a conventional \emph{workflow baseline}---a hand-crafted pipeline composed of instruction generation, rough-cut (visual clip composition), and fine-cut (TTS synthesis and post-production effects) stages executed in a fixed order. Such rigid execution prevents adaptation to diverse user-specific generation requirements, motivating the agentic design of VGAs.

As shown in Table~\ref{tab:consistency_success}, VGAs consistently outperform the baseline across all metrics. The gains stem from two capabilities: \emph{reasoning}, enabled by a hierarchically structured end-to-end framework that supports coherent cross-modal planning; and \emph{reflection}, which improves output quality through iterative self-correction and re-planning, capped at two iterations to maintain inference latency comparable to the workflow baseline while achieving substantial quality gains.





\vspace{2pt}
\noindent\textbf{Reward Contribution Analysis.}
We analyze the contribution of individual reward components under our synergistic cross-domain rewards. Since user feedback serves as the primary objective and is always retained, we focus the ablation on the two constraint-side rewards: video quality and interest alignment.

As shown in Table~\ref{tab:reward_abs}, each video-quality sub-reward is evaluated on its own dedicated test set, with \textit{Base} denoting the policy trained without any reward optimization. Each sub-reward---visual fidelity, audio alignment, and effect enhancement---independently improves the Automated Win Rate over the base, and jointly optimizing all three yields the strongest result, confirming the necessity of balancing all three perceptual aspects.

Building on the quality rewards, incorporating the interest alignment reward further lifts the Interest Alignment Score (0.707 $\rightarrow$ 0.828), indicating substantially stronger consistency between generated content and user interest.

Overall, these results show that the quality and alignment rewards play complementary roles---the former safeguards perceptual fidelity while the latter enforces semantic relevance---and their joint optimization, anchored by the primary user-feedback objective, produces a more robust and user-aligned video generation policy.

\subsubsection{Qualitative Analysis of Personalized Video Generation}
Figure~\ref{fig:e2e_overview} illustrates the end-to-end pipeline of our RaG framework, where user interests are directly transformed into video generation.

Given a representative user profile (female, 25--34) interested in young-mother lifestyle, baby care, household essentials, and value-oriented shopping, the Generative Recommendation Model (GRM) first infers Disentangled SIDs (D-SIDs) that jointly capture content and creative preferences. These D-SIDs are mapped into structured embeddings and, in this advertising scenario, further enriched with the optional metadata factor $D_{\text{meta}}$ encoding product information and marketing topics. Conditioned on these representations, the Instruction Model (IM) produces shot-level production instructions, which are subsequently executed by the Video Generation Agents (VGAs) to coordinate visual, audio, and effect generation. The resulting video achieves high quality with strong alignment to user interests.

%% file: kdd/2RelatedWork.tex
\section{Related Work}
\vspace{2pt}
\noindent\textbf{Retrieval-Based Recommendation.}
Traditional recommendation systems~\cite{covington2016deep, cheng2016wide, zhou2018din, zhu2018learning, naumov2019dlrm, jia2025learn} follow a retrieve-and-rank paradigm over a fixed pool of pre-produced items.

Recent advances explore generative recommendation by modeling item IDs as discrete tokens and formulating recommendation as next-token prediction over Semantic IDs~\cite{zheng2024lcrec, wang2024letter, yin2025ttds}. To meet industrial latency constraints, efficient architectures have been further proposed for large-scale SID prediction~\cite{deng2025onerec, xue2026gr4ad, zhou2025onerecv2technicalreport}.

However, these methods still rely on retrieving from a static content pool conditioned on predicted tokens, leaving recommendation and content creation fundamentally decoupled and preventing end-to-end optimization.

\vspace{2pt}
\noindent\textbf{Personalized AI-Generated Content.}
Recent progress in generative models has motivated a shift from retrieving pre-produced content to generating personalized content conditioned on user preferences. Early efforts explore preference-guided LLM generation or conditioning diffusion models on user signals for image synthesis~\cite{wang2023generec, yang2024cg4ctr, shen2024pmg}.

These ideas have been extended to richer modalities, including personalized advertising text generation~\cite{chen2025hllm} and dialogue-based preference elicitation for visual content generation~\cite{wang2025pcg}. More recently, NextAds~\cite{xu2026nextads} studies personalized video advertising by conditioning generation on observed user preferences, but focuses primarily on the generation module without modeling an end-to-end pipeline from user interest representation to controllable video production, and does not consider industrial deployment efficiency and cost.

Overall, existing approaches mostly treat user interest modeling and controllable content generation as separate tasks, leaving room for a unified framework that jointly optimizes both within a closed-loop industrial system.

%% file: kdd/6Conclusion.tex
\section{Conclusion}


We propose \textbf{Recommendation-as-Generation} (\textbf{RaG}), a unified paradigm that shifts recommendation toward generation-driven personalization. RaG bridges user interest modeling and controllable video generation through Disentangled Semantic IDs as a shared interface, scalable Video Generation Agents for industrial deployment, and Synergistic Cross-Domain Reward Learning for closed-loop optimization. Online A/B tests show that RaG consistently improves ad revenue over strong commercial baselines, demonstrating the effectiveness of the proposed paradigm in real-world industrial settings.

RaG currently serves nearline rather than in real time, with VGAs being the dominant latency bottleneck. Future work will fold the Instruction Model into VGAs for a tighter generation path, and further accelerate VGAs through stronger model distillation and inference optimization, moving toward on-the-fly personalized generation.

%% file: AppendixText/Appendix_full.tex
\section{Runtime Analysis of RaG Modules}
\label{app:runtime}

\begin{table}[t]
\centering
\small
\setlength{\tabcolsep}{4pt}
\renewcommand{\arraystretch}{1.2}

\caption{\small Inference efficiency comparison across modules of RaG system.}
\label{tab:efficiency}

\begin{tabular}{lcccc}
\toprule
\textbf{Component} 
& \shortstack{\textbf{D-SIDs}\\{(Nearline)}} 
& \shortstack{\textbf{GRM}\\{(Online)}} 
& \shortstack{\textbf{IM}\\{(Nearline)}} 
& \shortstack{\textbf{VGAs}\\{(Nearline)}} \\
\midrule

\textbf{Latency} 
& $\sim$4s 
& $\sim$100ms 
& $\sim$2.5s
& $\sim$180s \\
\bottomrule
\end{tabular}
\end{table}

We deploy the proposed RaG system in Kuaishou’s large-scale advertising and recommendation infrastructure, serving over 400 million users. To better understand system efficiency, we conduct a runtime analysis of each core component under both online and nearline deployment settings. Specifically, the Generative Recommendation Model (GRM) operates in an online serving regime for real-time recommendation, while Disentangled Semantic IDs (D-SIDs), the Instruction Model (IM), and the Video Generation Agents (VGAs) are executed in a nearline pipeline due to their higher computational cost and generation latency. All latency numbers reported in Table~\ref{tab:efficiency} are measured under live production traffic, reflecting the actual serving conditions.

\section{Evaluation Metrics for Video Quality and Interest Alignment}
\label{app:evaluation}

To provide a fair and unbiased evaluation of the proposed Video Generation Agents (VGAs), we adopt an evaluation protocol that is fully decoupled from the reward functions used during training. Specifically, we assess generated videos from three complementary perspectives: \emph{instruction-level interest alignment score}, \emph{automated multi-dimensional quality evaluation}, and \emph{human preference assessment}. All evaluation scores are computed using external judges or human annotators, ensuring that the reported results reflect generalization quality rather than optimization-specific reward fitting.

\vspace{2pt}
\noindent\textbf{Interest Alignment Score.}
We first evaluate whether generated videos faithfully follow their corresponding video production instructions derived from interest SIDs. To mitigate single-judge bias, we employ an ensemble of three state-of-the-art multimodal evaluators---GPT-5.1~\cite{gpt5_1_report}, Gemini-2.5~Pro~\cite{comanici2025gemini}, and Claude-4.5~Sonnet~\cite{claude45sonnet}---each independently scoring the same benchmark of 1{,}000 multi-category video instances under the identical protocol defined in Box~\ref{box:prompt_alignment}. Each generated video is rated along five dimensions---semantic consistency, attribute accuracy, thematic alignment, completeness, and narrative coherence---producing a continuous alignment score in $[0,1]$. We report the per-instance average across the three judges as the final Interest Alignment Score.

\vspace{2pt}
\noindent\textbf{Automated Quality Evaluation.}
Beyond semantic alignment, we further evaluate the overall production quality of generated videos using the same three-judge ensemble---GPT-5.1~\cite{gpt5_1_report}, Gemini-2.5~Pro~\cite{comanici2025gemini}, and Claude-4.5~Sonnet~\cite{claude45sonnet}---under the identical protocol defined in Box~\ref{box:prompt_quality}. Each judge independently rates every video along four aspects: (1) instruction attractiveness, measuring hook quality, pacing, and call-to-action effectiveness; (2) BGM compatibility, evaluating music-tone consistency and beat synchronization; (3) SFX and sticker design quality, assessing visual effects and subtitle design; and (4) instruction--visual alignment, measuring the consistency between visual progression and instruction semantics. Per-instance scores are averaged across the three judges, yielding two metrics: a normalized \textbf{Automated Score} in $[0,1]$ and an \textbf{Automated Win Rate} under the Good-Same-Bad (GSB) setting.

\vspace{2pt}
\noindent\textbf{Human Preference Assessment.}
To further validate real-world perceptual quality and user preference alignment, we conduct a human evaluation study with 20 annotators from diverse professional backgrounds---including algorithm engineers, product managers, and advertising clients---covering both algorithm-side and business-side perspectives to reduce single-role bias. Each annotator performs 50 pairwise comparisons between generated videos and baseline results under a Good-Same-Bad (GSB) protocol, yielding 1{,}000 pairwise judgments in total. All comparisons are presented in a blind, randomized order, with each video pair independently evaluated by at least three annotators to mitigate individual subjectivity; we report the majority-vote outcome as the \textbf{User Study Win Rate}.

\begin{promptbox}[after skip=3pt]{Interest Alignment Prompt}{prompt_alignment}

\textbf{Role}

You are an expert evaluator assessing the alignment between the video production instructions and the corresponding generated video. Focus only on semantic and creative consistency, while ignoring production quality (e.g., resolution, smoothness, or visual artifacts).

\medskip
\textbf{Inputs}

\begin{itemize}[leftmargin=*]
    \item \textbf{Instructions}: \{instructions\}
    \item \textbf{Video}: \{video\}
\end{itemize}

\medskip
\textbf{Task}

Evaluate the video along the following five dimensions, each scored in $[0,1]$.

\medskip
\textbf{[A1] Content Fidelity}
\begin{itemize}[leftmargin=*]
    \item Subjects, actions, and scenes match the instruction.
\end{itemize}

\textbf{[A2] Attribute Accuracy}
\begin{itemize}[leftmargin=*]
    \item Visual attributes and spatial-temporal relationships are correctly represented.
\end{itemize}

\textbf{[A3] Intent \& Theme Alignment}
\begin{itemize}[leftmargin=*]
    \item Creative intent, mood, and stylistic cues align with the instruction.
\end{itemize}

\textbf{[A4] Completeness}
\begin{itemize}[leftmargin=*]
    \item All key elements are included without hallucinated content.
\end{itemize}

\textbf{[A5] Narrative Coherence}
\begin{itemize}[leftmargin=*]
    \item The temporal progression and story flow remain coherent.
\end{itemize}

\medskip
\textbf{Scoring Scale}

\begin{itemize}[leftmargin=*]
    \item 0.9--1.0: Perfect alignment
    \item 0.7--0.9: Minor deviations
    \item 0.5--0.7: Moderate deviations
    \item 0.3--0.5: Major missing elements
    \item 0.0--0.3: Largely unrelated
\end{itemize}

\medskip
\textbf{Output Format}

\begin{verbatim}
{
  "content_fidelity": {"score": 0.0},
  "attribute_accuracy": {"score": 0.0},
  "intent_theme_alignment": {"score": 0.0},
  "completeness": {"score": 0.0},
  "narrative_coherence": {"score": 0.0},
  "overall_alignment_score": 0.0
}
\end{verbatim}

\textbf{Rules}

Scores should be continuous in $[0,1]$. The overall score is a holistic judgment rather than the arithmetic mean of sub-scores. Do not consider production quality unless explicitly required by the instruction.
\end{promptbox}

\begin{promptbox}[before skip=2pt]{Video Quality Assessment Prompt}{prompt_quality}

\textbf{Role}

You are a professional short-video advertising evaluator assessing video quality from editing and audio-visual perspectives.

\medskip
\textbf{Task}

Evaluate the video across four dimensions with a total score of 100 points.

\medskip
\textbf{[D1] Instruction Attractiveness (25)}
\begin{itemize}[leftmargin=*]
    \item Hook quality: pain-point, suspense, benefit-first, or contrast design.
    \item Pacing and structure: logical progression without redundant segments.
    \item CTA effectiveness: clarity, urgency, and consistency with opening intent.
\end{itemize}

\textbf{[D2] BGM Compatibility (25)}
\begin{itemize}[leftmargin=*]
    \item Mood and tempo alignment with video content.
    \item Synchronization between cuts and music beats.
    \item Balanced audio volume and speech clarity.
\end{itemize}

\textbf{[D3] SFX \& Sticker Design (25)}
\begin{itemize}[leftmargin=*]
    \item Effectiveness of transition, emphasis, ambient, and emotional SFX.
    \item Consistency and readability of subtitles, tags, arrows, and motion effects.
\end{itemize}

\textbf{[D4] Instruction--Visual Alignment (25)}
\begin{itemize}[leftmargin=*]
    \item Consistency between instruction keywords and visual content.
    \item Narrative flow and temporal coherence.
    \item Absence of visual-information gaps or dead-air segments.
\end{itemize}

\medskip
\textbf{Output Format}

\begin{verbatim}
{
  "instruction_attractiveness": {"total": 0},
  "bgm_compatibility": {"total": 0},
  "sfx_sticker_design": {"total": 0},
  "instruction_visual_alignment": {"total": 0},
  "final_summary": {
    "total_score": 0,
    "grade": "S/A/B/C/D"
  }
}
\end{verbatim}

\textbf{Rules}

The total score ranges from 0 to 100. Grades are defined as:
S (90+), A (75+), B (60+), C (45+), and D ($<$45). 
All evaluations should be evidence-based and supported by specific visual or temporal observations.

\end{promptbox}

\section{Details of Generative Recommendation Model}
\label{app:GRM}

Our Generative Recommendation Model (GRM) follows an architecture similar to GR4AD~\cite{xue2026gr4ad}. 

In training, the model takes two inputs: (1) user context $\mathbf{C}$, consisting of static profile features $\mathcal{F}_{\text{prof}}$ (e.g., age, gender, region, device type) and multi-granularity behavior sequences $\mathcal{F}_{\text{seq}}$, where each interaction is encoded via a sparse embedding table into latent interest tokens; and (2) prefix SID sequence {\small $(BOS, s_{\text{content}}^1, s_{\text{content}}^2, s_{\text{creative}}^1,\\ s_{\text{creative}}^2)$} derived from the target item. We formulate GRM as an autoregressive sequence modeling problem, predicting the full SID sequence conditioned on the prefix, where the generated SIDs serve as a discrete representation of user interests, optimized via token-level cross-entropy loss, followed by reinforcement learning fine-tuning with constrained GDPO to further improve user feedback and interest alignment.

Concretely, each SID token is retrieved from a sparse embedding table and projected into a 768-dimensional latent space, then processed by a 7-layer Transformer decoder (LazyDecoder) with hidden size 768, FFN size 3072, 12 attention heads, and vocabulary size 8192. FlashAttention~\cite{dao2023flashattention2} is adopted for efficient computation. We train the model on 8 GPUs with batch size 8192 using Adam ($lr = 1\times10^{-4}$). During inference, beam search (beam size 512) is used, achieving 130 QPS throughput.